\documentclass[aps,prd,showpacs,twocolumn,floatfix]{revtex4}
\usepackage{graphicx}
\usepackage{amsfonts}
\usepackage{amssymb}
\usepackage{amsmath}
\usepackage{flafter}
\usepackage{epstopdf}

\begin{document}

\title{Entanglement Entropy of the Early Universe in Generalized Chaplygin Gas Model}




\date{\today} 
\author{Pisin Chen$^{1,2,3}$}
\email{chen@slac.stanford.edu, pisinchen@phys.ntu.edu.tw}
\author{Yuezhen Niu$^4$}
\email{yuezhenniu@gmail.com}
\affiliation{1. Department of Physics and Graduate Institute of Astrophysics, National Taiwan University, Taipei, Taiwan 10617}
\affiliation{2. Leung Center for Cosmology and Particle Astrophysics (LeCosPA), National Taiwan University, Taipei, Taiwan, 10617}
\affiliation{3. Kavli Institute for Particle Astrophysics and Cosmology, SLAC National Accelerator Laboratory, Menlo Park, CA 94025, U.S.A.}
\affiliation{4. School of Physics, Peking University, Beijing 100871, China}

\begin{abstract}
We provide an explicit calculation of the evolution of the cosmic entanglement entropy in the early universe before the matter dominant era. This is made possible by invoking the generalized Chaplygin gas (GCG) model, which has the advantage of preserving unitarity and providing a smooth transition between the inflation epoch and the radiation dominant era. The dynamics of the universe is described by the quantization in the minisuperspace of the GCG model, following the prescription proposed by Wheeler and DeWitt. Two sources of contribution to the cosmic entanglement entropy are considered:
one from the homogeneous background where the observable and the unobservable regions 
of the universe are entangled and the other from the inhomogeneous cosmological perturbations where different modes are entangled.
We find that the homogeneous contribution grows exponentially at the very beginning of the inflation, but decreases during the radiation dominant era. Conversely, that from the cosmological perturbation is found to decrease at first and then increase after reaching a minimum value. The net result is that the total entanglement entropy reaches a minimum at an early stage of the inflation and then increases throughout most of the inflation and the entire radiation dominant era.
\vspace{3mm}


\end{abstract}

\maketitle
\section{Introduction}

An increasing amount of observational evidence supports the notion that the early universe has undergone an epoch of inflation. However we are still far from understanding the underlying assumptions and resolving some of its most crucial issues. One of the open questions about inflation is the entropy problem\cite{Penrose1979}\cite{Wald2006}\cite{Sean2010}, that is, why did the universe start from an extremely low entropy initial state, which in turn gave rise to the arrow of time? Since the initial entropy evaluates the probability
of spontaneous formation of a homogeneous domain of an
inflationary universe, the very low initial entropy implies the unnaturally small probability for our universe to evolve to the current state through inflation. Tantamount work has shed lights on possible solutions to this problem. For example one approach is the spontaneous eternal inflation where inflation can occur both forward and backward in time\cite{Sean2010}. This has also been treated in string theory in the form of bubble cosmology\cite{Coleman1980} where universes that realize the complete set of string vacuum solutions were constructed \cite{McInnes2007}\cite{McInnes2008}. More recently the tripartite partition of a physical system\cite{Tegmark2011} was introduced where the entropy was found to decrease dramatically after inflation as a means to solve the problem.

We note, however, that the definitions of the entropy budget are hugely different in these various approaches.  For example at late times when local thermal equilibrium can be reached in the universe, the notion of thermal entropy is applicable in issues related to cosmological structures such as galaxies or clusters. For strongly gravitating systems such as black holes, holographic entropy\cite{Takayanagi2006} satisfying the Beckenstein bound is used and most works approximate the total entropy of the universe as the sum of that of all black holes\cite{Kephart2003}\cite{Sean2004}\cite{Frampton2009}; in string theory\cite{McInnes2007} the Weyl curvature hypothesis\cite{Penrose1979} is often assumed which suggests that the arrow of time is due to the vanishing of the Weyl tensor that effectively acts as the entropy; for a universe dominated by ultrarelativistic matter, the total entropy coincides with the total number of particles in the universe\cite{Linde1999}. Apart from this lack of conformity in the definition of entropy when solving the entropy problem, all previous work only evaluated the entropy qualitatively due largely to the lack of a working model to track the entropy evolution in the early universe.

In principle the thermal entropy in a purely gravitating system is ill-defined for lack of a universal temperature. Also, statistical mechanics is not directly applicable due to the long-range nature of gravitational force. Self-gravitating
systems thus possess unusual features such as the absence of the global entropy maxima, different from conventional
thermodynamical systems\cite{Taruya2008}. Even so, the loss of information in a quantum gravitating system can still be evaluated by von Neumann entropy, which measures the degree of bi-partite entanglement. Since the early universe can be described as a quantum system without strong decoherence thereafter, it is justifiable to invoke von Neumann entropy in finding the evolution of the  total entropy of the universe, which is a dynamical quantum system far from thermal equilibrium. We will thus use the entanglement entropy instead of entropy in the following discussion to help clarify our setup and to quantify the amount of information in a quantum system.

In order to track the entanglement entropy of the universe, a specific model is needed. Most existing theories about the early universe necessitate the reheating process to connect the inflation to the subsequent radiation dominant era\cite{Linde2007}. However typical reheating mechanism such as the decay of the scalar field is not adequate to render tangible calculations of the entanglement entropy. As a remedy, we resort to an alternative approach, i.e., the generalized Chaplygin gas model (GCG), which bridges the inflation epoch and the radiation dominant era via a single equation of state and therefore a smooth transition. Though a phenomenological theory, GCG reproduces well the CMB power spectrum\cite{Lopez2011}. Although we invoke GCG so as to accomplish tangible calculations on entropy, the method developed in this work is general and can be used in arbitrary set up. The underlying effective Lagrangian of the theory, which is consistent throughout the whole period, yields a solvable Wheeler-DeWitt equation and the background evolution, both of which are crucial for the evaluation of entanglement entropy.

In Sect.\ref{chap}, we briefly introduced the GCG inflation model used to approximate the evolution of the early universe. This is the best candidate model so far in regards to providing a manageable framework for a specific calculation and the ability to reconstruct the existing observational data and make important predictions\cite{Lopez2011}. We expand the detailed method of finding the entanglement entropy in Sect.\ref{WDW} and \ref{pertSect} which make up two parts of the whole entanglement entropy: the homogeneous part and the inhomogeneous part. Numerical results are presented as figures followed with brief analysis in discussion where a decrease in homogeneous contribution is seen in radiation dominated era and a saturation to a lower bound appears in perturbation entanglement entropy.

\section{Generalized Chaplygin Gas\label{chap}}

When applied to cosmology, the Chaplygin gas\cite{Lopez2011} models the change of cosmic content by regulating the equation of state of the background
fluid instead of the form of the potential. It was first suggested by Kamenshchik\cite{Kamenshchik2002} in an attempt to smoothly interpolate the de Sitter phase and the radiation dominant era without ad hoc assumptions. Generalized Chaplygin gas models (GCG) were subsequently suggested\cite{Bento2002}\cite{Bertolami2004}\cite{Lopez2010}, some of which managed to unify dark energy and dark matter\cite{Bilic2002}. GCG was also
introduced, within the framework of FRW cosmology, as an exotic
background fluid to admit supersymmetric generalization\cite{Jackiw2001}. The equation of
state of GCG is given by\cite{Lopez2011}:
\begin{equation}
\rho=\Big(\frac{A}{a^{1+\beta}}+\frac{B}{a^{4(1+\alpha)}}\Big)^{\frac{1}{1+\alpha}}.
\end{equation}
From the conservation of energy the pressure for the
Chaplygin gas is
\begin{equation}
p= 1/3\rho + \frac{1 + \beta -4(1 + \alpha)}{ 3(1 + \alpha)}
\Big(\rho-\frac{B}{ a^{4(1+\alpha)}}\rho^{-\alpha}\Big).
\end{equation}
One can therefore find the underlying minimal coupling scaler field
$\phi$ and its potential $V(\phi)$ through the relations
\begin{equation}
\rho=\frac{\dot{\phi}^2}{2a^2}+V(\phi),\quad\quad\\
p=\frac{\dot{\phi}^2}{2a^2}-V(\phi).
\end{equation}
We require that the inflation takes place before radiation dominant era, which leads to the restriction:
\begin{equation}
\frac{B}{4(1+\alpha)}\ll\frac{A}{1+\beta}.\label{restric1}
\end{equation}
Two more conditions should be imposed
on $\alpha$ and $\beta$ to have a model that interpolates between
an early inflationary phase of the type of quintessence
and a subsequent radiation dominant phase: (i) the
energy density must induce a period of inflation
and (ii) the inflation should not cause a
super-inflationary expansion; i.e. $0< \dot{H}$. That is, a
super-accelerating phase of the universe, where the
energy density grows as the universe expands, should not occur. By
combining these two ansatz with the above inequality (Eq(\ref{restric1})), we
can easily deduce that the set of allowed values of $\alpha$ and
$\beta$ satisfy
\begin{eqnarray}
1+\alpha<0,\\ \nonumber
1+\beta<0,\\ \nonumber
2(1+\alpha)<1+\beta.\label{restric2}
\end{eqnarray}

The statistical feature of the Chaplygin gas can be deduced from the
first law of thermodynamic at much later time when thermal temperature can be well-defined:
\begin{equation}
T\mathrm{d}S=\mathrm{d}(\rho(T) V)+p(T)\mathrm{d}V,
\end{equation}
and its equation of state,
\begin{equation}
p=w \rho.
\end{equation}
The entropy is then
\begin{equation}
S=\frac{\rho+p}{T}.
\end{equation}
Let the holographic entropy be $S_A=\pi r_A^2/G$. The evolution of the total entropy can then be parameterized as
\begin{widetext}
\begin{equation}
\dot{S}+\dot{S_A}=3H\left[\frac{1+3w}{2}S_0\Big(\frac{a}{a_0}\Big)^{\frac{3(1+3w)}{2}}+(1+w)S_{A0}\Big(\frac{a}{a_0}\Big)^{3(1+w)}\right].
\end{equation}
\end{widetext}
Since the early universe can be treated as a quantum system without significant decoherence afterwards, thermal entropy should be replaced by the von Neumann entropy, which measures the degree of quantum entanglement. Without relying on further assumption of the area-law behavior of the holographic entanglement entropy, we try to find the total entropy budget by resorting purely to the von Neumann definition.

Once a quantum system is entangled, the wave-function of the system under consideration cannot be factorized. From the observational point of view, this means that we fail to access the entire phase space of the system since any single measurement will destroy the entangled state into a factorized state so as to conceal its entangled counterpart, which is exactly the process of decoherence. As a result, the degree of entanglement measures the loss of information, which is equivalent to the von Neumann entropy in a quantum system.
 
As we know, entanglement entropy from the von Neumann measurement counts the degree of bipartite entanglement and the result is dependent on the division of the system. Even for a system which is pure, the entanglement entropy of one subsystem can measure its degree of entanglement with the other subsystems. Multipartite entanglement entropy is thus needed to evaluate the overall degree of entanglement between all possible pairs of subsystems. There are definitions of multipartite entanglement entropy such as the global entanglement that maximizes the fidelity between a quantum state and a separable state\cite{Wei2003}. This definition, however, is difficult to employ for general systems.

We suggest that one reasonable way to quantify the total entanglement entropy of a quantum system is to sum up von Neumann entanglement entropies of all subsystems. A common approach to estimating the total entropy of the late time universe\cite{Kephart2003}, for example, is by adding the entropy of each black hole, which follows this same principle. The only difference is that this approach only considers the entanglement entropy between an individual black hole and the rest of the universe, whereas we here propose that the entanglement entropy between every pair of conceivable subsystems should be accounted in order to find the total entanglement of the whole quantum system. The value of multi-partite entanglement entropy in our definition can be infinitely large. However the absolute value of the total entanglement entropy may be less essential than its change. By focusing solely on the change of the entanglement entropy, our method can serve to unveil the dynamical behavior of the entanglement of the quantum system. One other virtue of our generalization lies in that the important features of the entanglement entropy found in previous studies such as subadditivity are well preserved.

Let us first define the early universe as a quantum system. The inflation in the early universe can be roughly divided into four stages: the onset of the inflation as a quantum gravitating system, the scaler field perturbations, the horizon exit and the radiation dominant era. Since the horizon re-entry happens after the radiation domination, the perturbations evolve during the radiation dominant era purely due to the expansion of the background metric. Since entanglement is not a statistical quantity, the entanglement entropy can be evaluated in systems far from equilibrium, which is the case for the highly dynamic universe during inflation. The evolution of the entanglement entropy in these different stages of inflation will be calculated explicitly in the following sections.

Suppose we can describe the state of a system completely by a set of discrete variables $O={O_1, O_2, ...}$. Once we find the variables of the system in certain values $O_i$, the system is in a pure state so that the conforming entanglement entropy is zero. However, if we know only the probability distribution $P_O$, then following von Neumann's definition, the entanglement entropy becomes $S=-\sum_O P_O \ln P_O$. If the variable is continuous, one takes the continuum limit and the entanglement entropy is in the form $S=-\int P(O) \ln P(O) \mathrm{d}O$\cite{Brandenberger1992}.

In order to evaluate the entanglement entropy based on this generalized multi-partite formalism, we need to determine the observables and their probability distributions. During the inflation epoch, it is reasonable to use the cosmic scale factor $a$ and the perturbation amplitude $\delta_k$ as our observables, since other physical quantities can either be determined by these two or are unobservable under current consideration. And there are two distinct contributions to the entanglement entropy of the early universe. We find the entanglement entropy contributed by $a$ in Section.\ref{WDW} by solving the wave-function from Wheeler-DeWitt (WDW) equation and then integrate it over $a$. This part of the entanglement entropy is due to the evolution of background metric and is thus a homogeneous contribution. On the other hand, the inhomogeneous part, accounted in Section.\ref{pertSect}, is contributed from the entanglement between two opposite modes $k$ and $-k$ of the adiabatic perturbation in a single-field Chaplygin gas. The total inhomogeneous entanglement entropy is then obtained by integrating over the entire mode spectrum.

\section{Entanglement from the wave-function of the early universe\label{WDW}}

Based on the canonical quantization approach to quantum gravity, WDW equation $H\Psi=0$\cite{DeWitt1967}\cite{DeWitt19672} determines the geometry of space-time of the minisuperspace of a FRW universe filled with Chaplygin gas. For a closed FRW model the minisuperspace Lagrangian is\cite{Monerat2007}\cite{Majumder2011}\cite{Lopez2005}
\begin{equation}
L=\frac{1}{2}\sqrt{-g}R+L_{CG}.
\end{equation}
The early universe is assumed to be homogeneous and isotropic that can be well approximated with the metric
\begin{equation}
\mathrm{d}s^2=-N^2\mathrm{d}t^2+a(t)^2\left[\frac{\mathrm{d}r^2}{1-r^2}+r^2 \mathrm{d}\Omega^2\right].
\end{equation}
We can then obtain the Lagrangian as
\begin{eqnarray}
\frac{1}{2}\sqrt{-g}R=-3\left(\frac{\dot{a}^2a}{N}-Na\right),\\
L_{CG}=-a^3N\rho_c=-a^3N\left(\frac{A}{a^{1+\beta}}+\frac{B}{a^{1+\alpha}}\right)^{1/(1+\alpha)}.
\end{eqnarray}
The momentum conjugate to the scale factor $a$, by definition, is
\begin{equation}
P_a=\frac{\partial(L_{CG}+\frac{1}{2}\sqrt{-g}R)}{\partial \dot{a}}=-\frac{6\dot{a}a}{N}.
\end{equation}
The Hamiltonian of the homogeneous and isotropic FRW universe with a positive curvature is thus
\begin{equation}
H=\frac{P_a^2}{12}+V_{\mathrm{eff}}(a).
\end{equation}
In our parameterization, the effective potential is
\begin{equation}
V_{\mathrm{eff}}=3a^2-\frac{a^4}{\pi}\left(\frac{A}{a^{1+\beta}}+\frac{B}{a^{1+\alpha}}\right)^{1/(1+\alpha)}.\label{veff}
\end{equation}
Clearly, the wave-function of the universe depends on the canonical variables $a$ and $\tau$: $\Psi(a,\tau)$, where $\tau$ is the conformal time. Following the standard procedure, we solve the wave-function by replacing $P_a$ with $i\frac{\partial}{\partial a}$ in the Hamiltonian, and the WDW equation becomes,
\begin{equation}
\left[{\frac{1}{12}\frac{\partial^2}{\partial a^2}-3a^2+\frac{a^4}{\pi}(\frac{A}{a^{1+\beta}}+\frac{B}{A^{4(1+\alpha)}})^{1/(1+\alpha)}}\right]\Psi(a,\tau)=0.
\end{equation}
The general solution to the wave-function can be expressed as
\begin{equation}
\Psi=C_1\psi_1+C_2\psi_2,
\end{equation}
where
\begin{equation}
\psi_i=\exp[-S^i(a)].
\end{equation}
$C_1$ and $C_2$ can be fixed by the boundary condition, and $S^i(a)$ satisfies the Hamilton-Jacobi equation:
\begin{equation}
\left(\frac{\mathrm{d} S_0^i(a)}{da}\right)^2=12V_0(a).
\end{equation}
The exact analytic solution for the effective potential in the form of Eq(\ref{veff}) is difficult to find. However, at very early times where the scale factor is very small, the potential can be approximated by
\begin{equation}
V(a)=A^{\frac{1}{\alpha +1}} \left(1-\frac{\beta +1}{6 (\alpha +1)}\right)a^{-\frac{\beta +1}{\alpha +1}}.
\end{equation}
Then the wave-function has a relatively simple solution:
\begin{widetext}
\begin{eqnarray}
\Psi=\exp\left[-2\frac{\sqrt{2} a(t) (\alpha +1) \sqrt{\frac{A^{\frac{1}{\alpha +1}}
   \left(6 \alpha -\beta +5\right) a(t)_1^{-\frac{\beta +1}{\alpha +1}}}{\alpha
   +1}}}{2 \alpha - \beta +1}+2\frac{\sqrt{2} a_0 (\alpha +1)
   \sqrt{\frac{A^{\frac{1}{\alpha +1}} (6 \alpha -\beta +5)
   a_0^{-\frac{\beta +1}{\alpha +1}}}{\alpha +1}}}{2 \alpha - \beta +1}\right].
\end{eqnarray}
\end{widetext}
We take $a_0=10^{-10}$ as our initial condition. The entropy can then be obtained by tracing over the even horizon:
\begin{eqnarray}
S&=&\mathrm{Tr}\left(\rho \mathrm{ln}\rho\right)\\
&=&\int_{a_{0}}^{a(\tau)}\Psi^\ast\Psi \mathrm{ln}(\Psi^\ast\Psi)\mathrm{d}a + S_0,
\end{eqnarray}
where the wave-function is already normalized and $S_0$ refers to the initial entropy. In this case, time variable is replaced by the scale factor and the background metric is determined by the equation of motion of the Chaplygin gas.

\begin{figure}[h]
\includegraphics[width=0.5\textwidth]{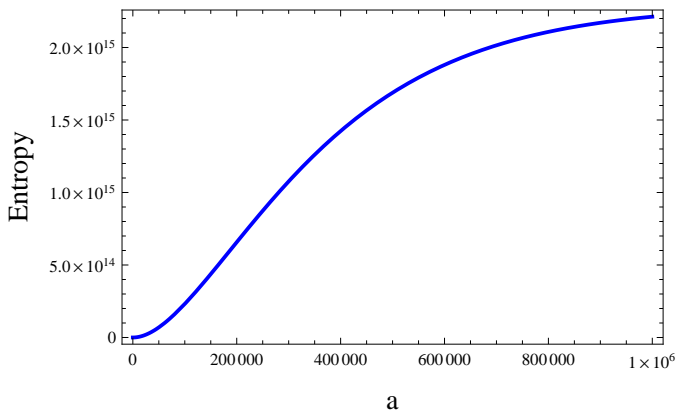}
\caption{The evolution of the entanglement entropy contributed from the entanglement between the observable universe and that outside the horizon from$a=10^{-9}$ to $a=10^6$, obtained by solving the WDW wave-function.}\label{initial}
\end{figure}

Carrying out the integration, the entropy is found to be
\begin{widetext}
\begin{equation}
S=f(\alpha, \beta, B, A) \Gamma \left(\frac{4
   \alpha -\beta +3}{2 \alpha -\beta +1},\frac{2 \sqrt{\frac{2}{3}}
   a(t)^{\frac{2 \alpha -\beta +1}{2 \alpha +2}} (\alpha +1)
   \sqrt{\frac{A^{\frac{1}{\alpha +1}} (6 \alpha -\beta +5)}{\alpha
   +1}}}{2 \alpha -\beta +1}\right)+S_0,
\end{equation}
\end{widetext}
where the coefficient $f(\alpha, \beta, B, A)$ is determined by $\alpha, \beta, B$ and A explicitly. The numerical result is shown in Fig.~\ref{initial}. We see that the entanglement entropy increases exponentially at the first stage of the cosmic inflation when the scale factor is also under exponential growth.

\begin{figure}[h]
\includegraphics[width=0.5\textwidth]{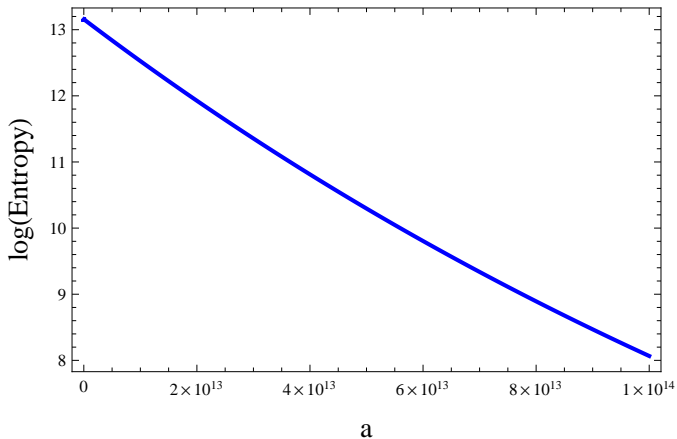}
\caption{The evolution of the entanglement entropy contributed from the entanglement between the observable universe and that outside the horizon in the radiation dominant era from $a=10^{12}$ to $a=10^{15}$.}\label{radiation}
\end{figure}

At a later time during the radiation dominant era, the effective potential can be approximated as
\begin{equation}
V(a)=\frac{1}{3}B^{1/(1+a)}a^{-4},
\end{equation}

and the wave-function is again solvable, and we find
\begin{equation}
\Psi=\exp\left[-2\left(\frac{1}{a_0}-\frac{1}{a(t)}\right)
   \sqrt{B^{\frac{1}{\alpha +1}}}\right].
\end{equation}
The corresponding entropy is found to be
\begin{widetext}   
\begin{eqnarray}
S=&&\frac{2}{3 a_0} \exp\left[-\frac{2 B^{\frac{1}{2(\alpha +1)}} }{\sqrt{3} a_0}\right]
\left(2 B^{\frac{1}{\alpha +1}}-\sqrt{3} a_0
   B^{\frac{1}{2(\alpha +1)}}\right) \text{Ei}\left(\frac{2
   B^{\frac{1}{2(\alpha +1)}}}{\sqrt{3} a(t)}\right)\\  \nonumber
   &&+ \sqrt{3}B^{\frac{1}{2(\alpha +1)}}
   \left(a_0 \exp\left[\frac{2
   B^{\frac{1}{2(\alpha +1)}}}{\sqrt{3} a_0}\right]-a(t)
   \exp\left[\frac{2 B^{\frac{1}{2(\alpha +1)}}}{\sqrt{3}
   a(t)}\right]\right)\\  \nonumber
   &&+\left(\sqrt{3} a_0
   B^{\frac{1}{2(\alpha +1)}}-2 B^{\frac{1}{\alpha +1}}\right)
   \text{Ei}\left(\frac{2 B^{\frac{1}{2(\alpha +1)}}}{\sqrt{3}
   a_0}\right).
\end{eqnarray}
\end{widetext}
This solution can be expressed in a simpler form:
\begin{equation}
S=-K_1 \text{Ei}\left(\frac{2
   B^{\frac{1}{2(\alpha +1)}}}{\sqrt{3} a(t)}\right)-K_2a(t)
   \exp\left[\frac{2 B^{\frac{1}{2(\alpha +1)}}}{\sqrt{3}
   a(t)}\right]+K_3,
\end{equation}
where $\text{Ei}(x)$ stands for the exponential integral function which increases monotonically for $x>0$, and $K_1$, $K_2$ and $K_3$ are positive constants. Interestingly, we find that the first derivative of the entanglement entropy is negative and thus it decreases monotonically. This conclusion is verified by a numerical computation (see Fig.~\ref{radiation})

However, since the total entanglement entropy contains other components apart from this homogeneous contribution from the wave-function of the background evolution, such as that from the cosmological perturbations, this decrease in entropy does not mean that the total entropy of the universe necessarily decreases.

\section{Entanglement entropy of cosmological perturbation\label{pertSect}}
The Mukhanov-Sasaki variable $\nu$ represents the linear combination of the inflaton
perturbations of the gravitational potential. The comoving curvature perturbation is then $\zeta(x, t)=\nu(x, t)/z(x, t)$, with $z(x, t)=a\sqrt{\epsilon}/4G$ and $\epsilon=-\mathrm{d}ln H/\mathrm{d}ln a$. The conjugate momentum of $\zeta$ is denoted as $\pi=\partial_{\mu}\zeta$. The Gaussian random state in a single scalar field inflation is then characterized by the covariant matrix between two modes $\textbf{k}$ and $\textbf{-k}$, which is related to the density matrix as\cite{Campo2008}
\begin{displaymath}
C=\frac{1}{e}\mathrm{Tr}(\rho{V, V^{\dagger}})=\left( \begin{array}{cc}
P_\zeta&P_{\zeta\pi}\\
P_{\zeta\pi}&P_\pi\\
\end{array} \right)\nonumber
\end{displaymath}

\begin{equation}
 \quad V=\left( \begin{array}{c}
\zeta_q\\
\pi_{-q}\
\end{array} \right).
\end{equation}
Each component can be identified as
\begin{eqnarray}
P_\zeta(q,t)=|\zeta_q|^2,\\
P_{\zeta,\pi}(q,t)=\frac{a^3 \epsilon}{4\pi G}\mathrm{Re}(\zeta_q \partial_t\zeta_q^\star),\\
P_{\pi}(q,t)=(\frac{a^3 \epsilon}{4\pi G})^2 |\partial_t\zeta_q|^2.
\end{eqnarray}
The corresponding entropy can then be found:
\begin{eqnarray}
S=2[(\bar{n}+1)\ln(\bar{n}+1)-\bar{n}\ln\bar{n})],\\
(\bar{n}+\frac{1}{2})^2=P_\zeta P_\pi-P_{\zeta\pi}^2.
\end{eqnarray}
The field $\nu(k)$ satisfies in the Fourier space
the equation
\begin{equation}
\ddot{\nu}+\frac{3\nu}{t}\dot{\nu}+\frac{q^2}{(\gamma t)^{2\nu}}\nu=0,\label{perturbation}
\end{equation}
which governs the evolution of $\zeta(k)$.

\begin{figure}[h]
\includegraphics[width=0.5\textwidth]{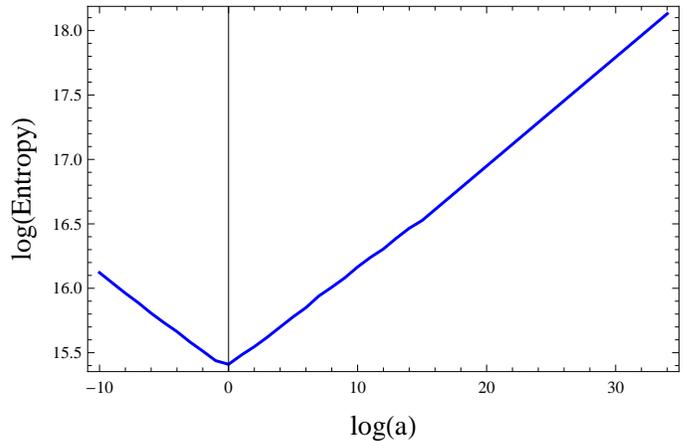}
\caption{Numerical result for the entanglement entropy contributed from the cosmological perturbation from the initial state of asymptotic Minkowski perturbation at $a=10^{-10}$ to a time well after the horizon exit for most modes at $a=10^{34}$.}\label{ppf}
\end{figure}

So the basic steps to calculate the entropy are: i) evolve the background metric dictated by Eq.~(\ref{cosmicscaler}); ii) solve the comoving curvature perturbation at time $t$ for different $k$ according to Eq.~(\ref{perturbation}); iii) integrate the total entropy over the momentum space following Eq.~(\ref{totalentropy}). The numerical results is shown in Fig.~\ref{ppf}.

Mind that though from the current observation, we cannot determine the exact starting time for the expected cosmological perturbation, a rough upper limit can be given by requiring the energy scale not to exceed the planck mass. The corresponding time for this limit is at approximately $a=10^{-35}$ much below our concerned value in calculation. 

One way to look into the entanglement behavior at much earlier time is by assuming the vacuum state as the asymptotic initial condition. Here, for example, the free vacuum is heuristically taken to be the Bunch-Davies vacuum. Inserting the background evolution into our GCG model\cite{Lopez2011}, we find
\begin{eqnarray}
a&=&\left[\frac{1+\beta}{2(1+\alpha)}A^{\frac{1}{2(1+\alpha)}}\sqrt{\frac{\kappa^2}{3}}t\right]^{\frac{2(1+\alpha)}{1+\beta}}\\ \nonumber
&=&(\gamma t)^\nu ,\label{cosmicscaler}
\end{eqnarray}
where $\kappa^2=8\pi G$, so that the entropy of the whole space under the assumption of Bounch-Davies vacuum is
\begin{widetext}
\begin{eqnarray}
S=\int_{q_{\mathrm{min}}}^{q_{\mathrm{max}}}\mathrm{d}q\cdot 2\Big[\Big(\frac{|\zeta_q^0|^2}{4\pi G\nu^3\gamma^\nu}t^{2-\nu}q^3+\frac{1}{2}\Big)\ln\Big(\frac{|\zeta_q^0|^2}{4\pi G\nu^3\gamma^\nu}t^{2-\nu}q^3+\frac{1}{2}\Big)\\ \nonumber
-\Big(\frac{|\zeta_q^0|^2}{4\pi G\nu^3\gamma^\nu}t^{2-\nu}q^3-\frac{1}{2}\Big)\ln\Big(\frac{|\zeta_q^0|^2}{4\pi G\nu^3\gamma^\nu}t^{2-\nu}q^3-\frac{1}{2}\Big)\Big],\label{totalentropy}
\end{eqnarray}
\end{widetext}
where $q_{\mathrm{max}}=2\pi/l_p$ and the particle horizon of the system, $L=\int_0^t dt/a(t)$, determines the ultraviolet cutoff $q_{\mathrm{min}}=2\pi/L$, which can also be expressed in terms of $t$ in our model as $q_{\mathrm{min}}=[2\pi\gamma^\nu/(1-\nu)]t$.

It is easy to prove that for $||\delta_q^0|^2t^{2-\nu}q_{\mathrm{min}}^3/(4\pi G \nu^3\gamma^{3\nu})|<1/2+1/e$, the entropy increases monotonically with time. Since the scalar perturbation reaches its asymptotic maximum at the end of the free scale perturbation, i.e.,  $|\delta_{\mathrm{max}}|=|\delta_{\mathrm{horizon exit}}|$, we can give a rough estimate of the time when entropy henceforth increases incessantly:
\begin{equation}
t\geq\left[\frac{(1+2e^{-1}) G \nu^3(1-\nu)}{|\delta_{\mathrm{max}}|^2}\right]^{1/3(\nu+1)}.
\end{equation}0000uuuudjfkdhfjd

\section{Discussion}
We have seen that the generalized Chaplygin gas model, though a phenomenological approach, may well provide a smooth transition between the inflation and the radiation dominant phases in the early universe. On the other hand, it fails to capture one essential feature common to most other inflation models, namely the post-inflation reheating process where copious particles are created. However, if as suggested\cite{Sean2010}\cite{Sean-Tam2010} that the evolution of the universe is unitary, GCG model which preserves unitarity may have the advantage over other inflation models in solving the entropy problem.

One should keep in mind that the results obtained in the above sections, in particular that shown in Fig.\ref{initial}, \ref{radiation}, and \ref{ppf}, only concern the variation instead of the absolute value of the total entanglement entropy. In our approach, we have evaluated the entropy evolution contributed from the homogeneous background and the inhomogeneous perturbation dynamics, from the very beginning of the inflation to the radiation dominant era. We found that the entanglement entropy contributed from the wave-function in the radiation dominant era undergoes a monotonic decrease. But the total entanglement entropy still increases during the same period due to the larger counteracting contribution from the curvature perturbation. However we found, as shown in Fig.\ref{ppf}, that at the much earlier time when the inflation has just started, the total entanglement entropy has indeed reached a minimum at one time. This interesting turnover happens at $a=1$ in our unit, which corresponds to $a=10^{-58}$ in the usual notation. This unexpected result underlies the possible existence of a lower bound in the entanglement entropy that conforms with the proven bound in a much simpler quantum system\cite{Verstraete2004}. Whether the existence of such a lower bound is universal remains to be tested.

\section{Acknowledgement}
We thank Mariam Bouhmadi-Lopez and Yen-Wei Liu for helpful discussions on the issues related to the generalized Chaplygin gas model. This research is supported by National Fund for Fostering Talents of Basic Science
(Grant Nos. J0630311, J0730316) in China, by Taiwan National Science Council under Project No. NSC
97-2112-M-002-026-MY3, and by US Department of Energy under Contract No. DE-AC03-
76SF00515. We also thank the support of the National Center for Theoretical Sciences
of Taiwan.


\begin{references}


\bibitem{Penrose1979}
Penrose, R. (1979). Singularities and time asymmetry. In S. W. Hawking,  W. Israel (Eds.), General relativity, an
Einstein centenary survey. Cambridge: Cambridge University Press.

\bibitem{Wald2006}
R. M. Wald
Modern Physics 37 (2006) 394C398

\bibitem{Sean2010}
S. Carroll, From Eternity to Here: The Quest for the
Ultimate Theory of Time (New York, Plume, 2010).



\bibitem{Coleman1980}
S. R. Coleman and Frank De Luccia
Phys. Rev. D {\bf21}, 3305 (1980).


\bibitem{McInnes2007}
B. McInnes
Nuclear Physics B {\bf782} 1 (2007)


\bibitem{McInnes2008}
B. McInnes
Phys. Rev. D {\bf77}, 123530 (2008)

\bibitem{Tegmark2011}
M. Tegmark
arXiv:1106.0007


\bibitem{Takayanagi2006}
S. Ryu, T. Takayanagi
JHEP 0608:045,2006

\bibitem{Kephart2003}
T. W. Kephart1 and Y. J. Ng
JCAP  {\bf11} 011 (2003)

\bibitem{Sean2004}
S. Carrall and J. Chen
Arxiv: hep-th/0410270



\bibitem{Frampton2009}
P. Frampton, S. D. H. Hsu, T. W. Kephart,
and David Reeb
arXiv:0801.1847.

\bibitem{Linde1999}
N. Kaloper, A. Linde, and R. Bousso
Phys. Rev. D {\bf59}, 043508 (1999)


\bibitem{Taruya2008}
A. Taruyaa and M. Sakagamib,
Physica A: Statistical Mechanics and its Applications
{\bf307}, Issues 1-2,(2002)

\bibitem{Linde2007}
A Linde - Inflationary Cosmology, 2007 - Springer


\bibitem{Lopez2011}
M Bouhmadi-Lopez, P Chen,Yen-Wei Liu
arXiv:[1104.0676]

\bibitem{Kamenshchik2002}
A Kamenshchik, U Moschella
Physics Letters B, 2001 - Elsevier

\bibitem{Bento2002}
M. C. Bento, O. Bertolami, and  A. A. Sen
Phys. Rev. D {\bf 66}, 043507 (2002)


\bibitem{Bertolami2004}
O. Bertolami, A. A. Sen, S. Sen, P. T. Silva,
Monthly Notices of the Royal Astronomical Society {\bf353} (2004)

\bibitem{Lopez2010}
M. Bouhmadi-Lopez, P. Frazo and A. B. Henriques
Phys. Rev. D {\bf81}, 063504 (2010)

\bibitem{Bilic2002}
N. Bilic, G.B. Tupper and R. D. Viollier
Physics Letters B, {\bf535}, (2002) - Elsevier

\bibitem{Jackiw2001}
Y. Bergner, R. Jackiw, Phys. Lett. A 284, 146 (2001)

\bibitem{Wei2003}
T.-C. Wei and P. M. Goldbart, Phys. Rev. A {\bf 68}, 042307 (2003).

\bibitem{Nishioka2009}
T. Nishioka, S. Ryu and T. Takayanagi
J. Phys. A: Math. Theor. {\bf 42}, 504008 (2009)

\bibitem{Brandenberger1992}
R. Brandenberger, V. Mukhanov, and T. Prokopec
Phys. Rev. Lett. {\bf69}, 3606 (1992)

\bibitem{DeWitt1967}
B S DeWitt
Phys. Rev. {\bf160}, 1113¨C1148 (1967)

\bibitem{DeWitt19672}
B S DeWitt
Phys. Rev. {\bf162}, 1195¨C1239 (1967)


\bibitem{Monerat2007}
G. A. Monerat
Phys. Rev. D {\bf76}, 024017 (2007).

\bibitem{Majumder2011}
B. Majumder
Physics Letters B, (2011).

\bibitem{Lopez2005}
M. B. Lopez and P. V. Moniz
Phys. Rev. D {\bf 71}, 063521 (2005)

\bibitem{Campo2008}
D. Campo and R. Parentani
Phys. Rev. D {\bf78}, 065044 (2008)



\bibitem{RAPHAEL}
R. Lamon and A. J. Wohr Phys. Rev. D {\bf 81}, 024 (2010)



\bibitem{Sean-Tam2010}
S. Carroll and H. Tam,
arXiv:1007.1417 [hep-th] (2010).


\bibitem{Verstraete2004}
F. Verstraete, M. Popp, and J. I. Cirac
Phys. Rev. Lett. {\bf92}, 027901 (2004)




\end{references}
\end{document}